# Lack of Resilience in Transportation Networks: Economic Implications

Margaret Kurth [1], William Kozlowski [2], Alexander Ganin [1,3], Avi Mersky [1], Billy Leung [2], Maksim Kitsak [4], and Igor Linkov [1]

## Abstract

Disruptions to transportation networks are inevitable. Currently, most mandated development-related transportation planning is intended to prepare for frequently occurring and observable disruptions while low probability events that have not yet materialized attract less attention. When road networks are not resilient, i.e., they do not recover rapidly from disruptions, these unpredictable events can cause significant delays that may not be proportional to the extent of the disruption. Enhancing resilience can help in mitigating consequences of disruptions but requires financial investment that is difficult to justify given that low probability event may not have materialized. This paper highlights economic implications of unmitigated random disruptions in urban road systems and makes the case for investment in transportation network resilience. We utilized a model of urban transportation network performance that quantifies resilience and demonstrated its integration with microeconomic transportation planning model REMI TranSight. The model was applied to 10 cities in the USA to calculate several economic indicators under baseline scenario where economic impact was assumed to be proportional to the magnitude of disruptive events and under a test scenario where the magnitude of disruption was used to calculate additional delays in transportation networks and these additional delays were explicitly integrated in REMI model. Results show that GDP losses suffered as a result of disruption may be far more significant in the case scenario and economic output is not necessarily back to normal in the year the following disruptive event. We thus conclude that support for investment decisions on network efficiency and resilience management should be based on a framework that utilizes resilience, quantified in terms that are compatible with standard practice, and scenarios to test the implications of topological attributes.

## Introduction

Transportation system investments are principally motivated by the goal of reducing delay (Belenky, 2011) and, by nature of modeling norms and associated metrics, projects are designed to do so by targeting improvements in efficiency. Current practice is to evaluate road performance with Level of Service (LoS), or similar measure of efficiency, during the worst traffic of an average day and when the whole road network is running as expected (*Highway Capacity Manual, 5th Edition*, 2010). The current norm for improving transportation networks and remedying the economic impact of delays is undertaken through an efficiency lens and to a growing extent, through management for specific threats. Emphasis on travel time and the monetary value of that duration allows prospective projects to enter the realm of cost-benefit analysis but carries the short-coming of representing only average conditions, not extremes.

---

[1] U.S. Army Corps of Engineers, Engineer Research and Development Center
[2] Regional Economic Models Inc.
[3] University of Virginia
[4] Northeastern University

The transportation field recognizes that variances in expected travel times have a cost, even when not incurred, as they need to be planned for by travelers. This is known as the Value of Reliability (U.S. Department of Transportation, 2016). While Value of Reliability is well researched and methodologies to estimate it do exist (Fosgerau and Karlström, 2010; Lam and Small, 2001), no standardized method has yet been adopted in the US (U.S. Department of Transportation, 2016). These costs may not be consistent across sectors and some workers may have more flexibility in where and when they work than others. These costs may also fluctuate with the amplitude of the delay, with small delays potentially being negligible, though more research is needed (Fosgerau et al., 2007; Mackie et al., 2003; U.S. Department of Transportation, 2016). A more standard metric used in planning is Value of Time (VOT), which is generally calculated on the assumption that variance in travel time from one scenario to another is certain and value is linearly calculated based on wage rates (U.S. Department of Transportation, 2016). Both of these methods fail to capture the costs of unpredictable delays, which cannot be accounted for in schedules and may have different costs than those associate with time passed in traffic.

The realm of possibilities between average traffic conditions and natural disasters is comprised of events that can have non-trivial disruptions to mobility, and yet are outside the realm of planning and analysis. These events may be unpredictable in space and time and unknown in nature to the point of being random and therefore, should be the focus of resilience inquiries here and elsewhere (Ganin et al., 2019, 2017). Whereas mitigating risk of disruption (i.e., strengthening key nodes and links in transportation networks) is appropriate for specific hazards, events that are highly uncertain in space and time challenge our ability to characterize vulnerability to them and implement effective risk mitigation measures. Cost concerns ensure that completely minimizing physical risk at one location may inherently limit our ability to reduce risk elsewhere. Similarly, hardening a transportation system against the risk of new types of disruptions, such as cyber-attacks on Intelligent Transportation Systems (ITS), is difficult because the potential range of risk is too vast to be effectively predicted (Ganin et al., 2019). Network-wide management is therefore more appropriate than location-specific solutions, where the objective is to keep people and goods flowing through the network in spite of disrupted parts of the network. An apt measure of a network's performance with respect to that objective is resilience. Resilience in transportation manifests is the ability to function in spite of disruption and/or recover function rapidly following disruption.

Across various contexts, there is a growing recognition that lack of resilience can have grave socioeconomic consequences, especially in the context of damaged interconnected infrastructure (Florin and Linkov, 2016). Such is the finding of a recent World Bank report on infrastructure as a key enabler of economies and the macroeconomic impacts of not being resilient (Hallegatte et al., 2019). Presidential Policy Directive 21 - Critical Infrastructure Security and Resilience (Obama, 2013) formalized the call to enhance the nation's critical infrastructure functioning and resilience by recognizing the importance of operable critical infrastructure, including transportation systems, one of 16 that are called out as vital to national economic security, public health, and safety ("Critical Infrastructure Sectors," 2019). The above synopsis of the current state of practice in transportation planning lends itself to the conclusion that advances in resilience research need to be integrated into planning norms to help account for uncertain events and emerging risks. Quantitative resilience modeling results can be used in tandem with efficiency-driven modeling efforts to conduct tradeoffs among multiple objectives in transportation network investment.

This paper aims to demonstrate that dovetailing resilience analysis with regional economic modeling can advance the methodological approach necessary for planning. Determining the extent of socioeconomic impacts due to disrupted infrastructure is actively being researched but importantly, the field is largely limited to forecasting or assessing impacts of specific disruptive events. For example, (Ham et al., 2005) assessed the anticipated economic implications if an earthquake were to occur in the New Madrid Seismic Zone in the U.S. Midwest and concluded that the ensuing disruption to U.S. commodity flow could pose significant threat to economic stability and recovery at the regional, national, and international scale. Transportation network resilience, according to the (Ham et al., 2005) context, refers the adaptability of commodity flow such that goods can be transported via multiple modes. Similarly, (Tatano and Tsuchiya, 2008) developed a spatial computable general equilibrium (CGE) model to estimate economic losses (e.g., changes in the cost of travel time) attributed to earthquake disruption of freight and passenger transportation flow. They used the 2004 Niigata-Chuetsu Japan earthquake as a case study, and regional economic losses were measured as a function of inter- and intraregional trade (Tatano and Tsuchiya, 2008). (Pelling et al., 2002) discuss how the 1995 Kobe, Japan earthquake increased transportation costs in the region by over 50% and increased the cost of goods in the region by 10%. Internationally, the disablement of the Kobe Port halted the import and export of goods. (Pelling et al., 2002) suggest that disasters possess an "inflationary potential" due to their capacity to affect the "production, distribution, marketing, and consumption" functions of markets. (Cho et al., 2015) show how resilience, given disruption to critical highway infrastructure, such as highway bridge and tunnel damage, can affect economic losses on a U.S. state-by-state and industry basis. They conclude that the states and industries that can adapt to disruption suffer the least economic loss and therefore redundancy in transportation networks can mitigate impacts. This conclusion is mirrored by (Worton, 2012), who suggested that resilience engineering should focus less on efficiency than the capacity for preparedness, recovery, and adaptation (Mattsson and Jenelius, 2015). Thus the shortcomings of existing studies on the economic implications of transportation disruptions are twofold: 1) resilience is often conflated with potential damages and 2) economic modeling is not underpinned by analysis of transportation network topology.

Resilience planning should use economic modeling that is based in transportation network analysis so that transportation costs are associated with travel time delays, as they are the cause of economic impact. Here we use REMI TranSight (*TranSight 4.2 User Guide*, 2018), an established economic modeling process that uses input-output, computable general equilibrium, econometric and economic geography methods to reinforce that lack of resilience of transportation networks to disruption can have non-trivial economic consequences via the delays that they cause to travel. Disruption-induced delay results generated by (Ganin et al., 2017), which fall significantly outside of the range of delay that is generally used in scenarios that underlie policy, are translated to economic outcomes by formulating them as inputs to REMI TranSight. This creates a process for quantifying the economic implications of resilience or lack thereof and progresses a planning approach that explicitly considers resilience. The method is demonstrated for ten U.S. cities. The methodology used in this study can be instrumental in the transition from current risk-based planning to true resilience planning, supported by economic analysis and subsequent selection of management alternatives.

# Methods: Integration of Transportation Network Research and Economic Modeling

This paper joins two independently developed and documented models to assess economic impacts of resilience in transportation networks: 1) (Ganin et al., 2017)'s network model for simulating traffic in urban road networks and calculating travel times associated with predictable disruptions (e.g., during pick commute) and random disruptions and accidents (e.g., natural disasters and major accidents) and 2) TranSight (*TranSight 4.2 User Guide*, 2018), a regional economic forecasting model oriented specifically for simulating the outcomes of changes in transportation systems.

## Network Model for Simulating Delays Associated with Disruptions

Effects of disruptions on an urban transportation infrastructure are quantified with the model proposed in (Ganin et al., 2017). Specifically, the model assesses travel delays for peak-hours private vehicle commuters in 40 urban areas in the continental U.S. Urban areas are densely developed and encompass residential, commercial, and other non-residential urban land uses ("2010 Census Urban and Rural Classification and Urban Area Criteria," 2018). An example of the Houston, TX urban area is given in Figure 1.

In order to study travel delays in each of the urban areas under normally functioning and disruption scenarios, (Ganin et al., 2017) first built transportation graphs comprised of intersections connected by roadways and then generated trips based on population assigned to each intersection. The population assignment was done with Voronoi tessellation and data from the U.S. Census Bureau. Trip distribution was accomplished with a modified gravity model and only privately-owned vehicle (POV) mode of travel was studied. Route assignment was done assuming free-flow speeds on all roadways. Next, based on traffic volumes on each roadway, the authors proposed a model to estimate the effective traffic speed. Using those congestion-based speeds, the authors evaluated the ensuing *total annualized travel time*. The model was subsequently calibrated to match measured data on annualized delay per a peak-hours auto commuter as given by the Urban Mobility Scorecard (Schrank et al., 2015). Modeled delays were calculated as the difference between travel times under congestion and travel times with free-flow speeds. The details of each step of the model development can be found in (Ganin et al., 2017).

To characterize the resilience of modeled cities, (Ganin et al., 2017) generated disruptive events on the transportation networks by disabling links and quantified resilience as additional delay resulting from these events. More specifically, they randomly selected a fixed fraction $\rho$ of network links that was made non-functional by reducing their free-flow speeds to 1 km/h. Links were selected at random, with probabilities proportional to their lengths to account for the fact that longer roads are more likely to be affected by adverse events. Then traffic was redistributed per the updated link free-flow travel times assuming the same origin-destination demand matrix. Additional delays that resulted from those events were evaluated as the annualized travel time difference in presence of a disruption and without one.

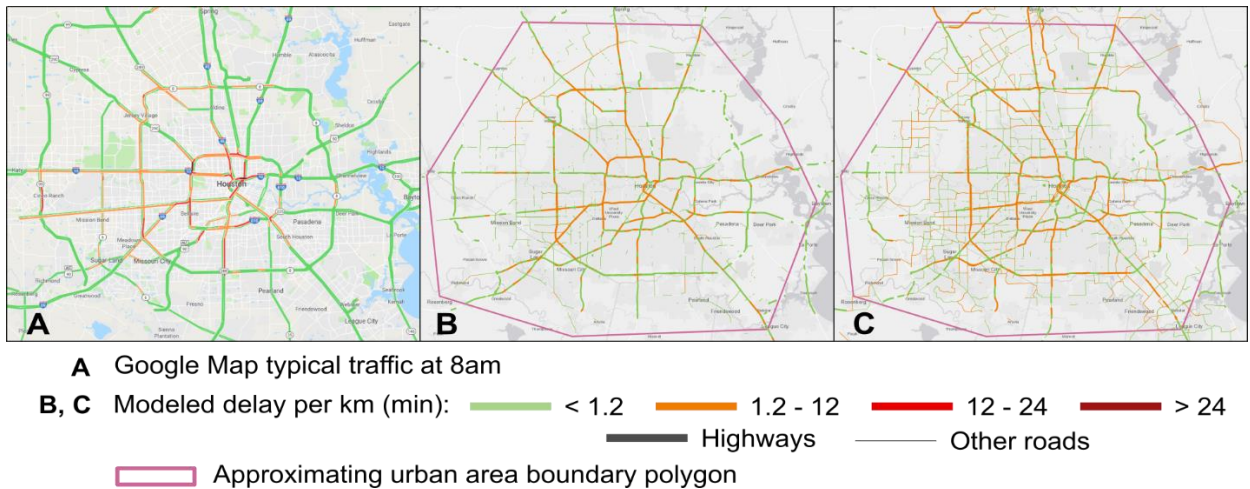

**Figure 1. Roadway networks of Houston, TX: (A) Congestion patterns at 8 am per Google Maps, (B) Modeled delays per kilometer of travel in normal conditions (B) and under 5% disruption (C). Purple line shows an approximation of the urban area boundary in panels (B) and (C).**

Economic Modeling for Insight into Impacts of Transportation Network Changes
TranSight (*TranSight 4.2 User Guide*, 2018) is designed to be used with transportation forecasting models to translate the outcomes of improvement measures into regional economic implications. In this case, instead of forecasting models, the transportation resilience model by (Ganin et al., 2017) was used to generate additional travel time (delay) that results from transportation system disruptions. For transportation studies, cost-savings, capital investment, and other financial and economic concerns associated with prospective infrastructure projects are related to the regional economy via changes to economic variables, which are called "policy variables" in the model. These policy variables represent the impact of travel time on individuals via, for example, spending on fuel and subsequently their disposable income and consumer spending, as well as on industries via, for example, extent to which labor demand is met and composite price of goods they send to market. All of the linkages between travel times and associated costs to the policy variables that are used in regional economic models are detailed in (*Model Equations*, 2017; *TranSight 4.2 User Guide*, 2018). TranSight can also account for the economic effects of changes to emissions, safety, and time saved as a result of transportation projects. These effects enter the economy as a change in the non-pecuniary amenity policy variable, which accounts for the desirability of an area as a place to live, independent from financial concerns such as wage levels.

Regional Economic Models Inc., the developer of TranSight and related products, maintains regional economic models for a wide variety of U.S. regions and states in order to support such studies. For a particular city or region, economic effects of transportation projects are forecasted in economic terms that include gross domestic product (GDP), employment, delivered price, commodity access, labor access, and relative cost of production. These are the outputs of PI+ Engine, a model that mixes techniques from Input-Output (I-O) and Computable General Equilibrium (CGE) modeling, as well as economic geography and econometric techniques (*Model Equations*, 2017). The transportation-economic modeling sequences (TranSight with PI+ Engine) typically supports the evaluation of alternative transportation projects and policies.

The inputs for TranSight from a transportation forecasting model are: the change in Vehicle Miles Traveled *VMT*, Vehicle Hours Traveled *VHT*, and vehicle trips attributable to improvement measures, such that changes to the transportation system will yield changes in figures such as average *VMT/VHT*, i.e., average travel velocity, or Trips/VHT, i.e., average delivery trips that can be made in a given amount of time. In TranSight, changes in velocity from the baseline scenario to the improvement (test) scenario is formulated as proportional to effect on *transportation cost* and changes in trips that can be made is formulated as proportional effect on *accessibility cost.* Transportation projects are presumed to have an effect on various economic variables, via changes in "effective distance", which functions to change travel time, or commuting time and expenses. Cost-savings due to reduced travel times accrue to industry firms in the model from reduced commuting and transportation costs, and increased access to markets.

## Connecting Transportation Network Resilience and Regional Economic Modeling

For the specified road network disruption severity $\rho$, quantifying the fraction of affected roadways, the network model (Ganin et al., 2017) calculates the resulting average annualized travel time $T(\rho)$ per a peak-hours commuter. The topological attributes that yield higher or lower $T(\rho)$ values are not called into question in this research.

We assume a linear relation between the transportation costs and the travel time, estimating the corresponding percent increase in transportation costs $c(\rho)$ as

$$c(\rho) = \Delta T(\rho) \big/ T(0) \qquad (1)$$

where $\Delta T(\rho) = T(\rho) - T(0)$.

For the purposes of this demonstration, and given the small sample size of cities studied, we rely heavily on changes in Gross Domestic Product (GDP) as an indication of economic impact and do not probe the mechanisms of the economy that underlie the overall impact. To quantify the effects of transportation cost increase on GDP we utilize the TranSight model, and then generate the relative change in GDP as function of disruption severity for the 10 cities of interest. Two scenarios are modelled:

1) the baseline scenario assumes that transportation cost increases are directly proportional to road disruption severity $\rho$:

$$c^0(\rho) \equiv \rho, \qquad (2)$$

2) the test scenario assumes that transportation cost increases are proportional not to the fraction of roads affected but to the additional travel time induced by disruption, see Equation 1.

## Results

Table 1 displays the percent increase in transportation cost computed for different cities at specific values of road network disruption severity $\rho$ and serves as the base for subsequent analysis of economic implications.

**Table 1. Transportation cost $c(\rho)$ increases by city that result from road network disruptions of severities varying from 1 to 5 percent. Shown in red are the transportation cost increase values exceeding 25%.**

| | | Fraction of Affected Roadways (Network Links), $\rho$ | | | | |
|---|---|---|---|---|---|---|
| | | 1% | 2% | 3% | 4% | 5% |
| Transportation Cost Increase, $c(\rho)$ | Atlanta | 4% | 10% | 16% | 23% | **33%** |
| | Detroit | 3% | 6% | 9% | 14% | 19% |
| | Houston | 5% | 11% | 16% | 24% | **32%** |
| | Jacksonville | 7% | 13% | 22% | **33%** | **44%** |
| | Los Angeles | 1% | 3% | 5% | 7% | 9% |
| | Miami | 4% | 9% | 13% | 18% | 23% |
| | Orlando | 4% | 9% | 14% | 20% | **26%** |
| | San Francisco | 9% | 20% | **34%** | **43%** | **51%** |
| | Seattle | 3% | 6% | 9% | 13% | 17% |
| | Tampa | 6% | 12% | 20% | **26%** | **37%** |

We observe that the same disruption results in different cost increases across multiple cities. Economic impacts of transportation disruption are most pronounced in San Francisco: here 5% disruption results in 51% increase in transportation costs, while similar disruption in Los Angeles results in only 9% increase. Jacksonville is the second most sensitive city considered, with 44% increase in cost.

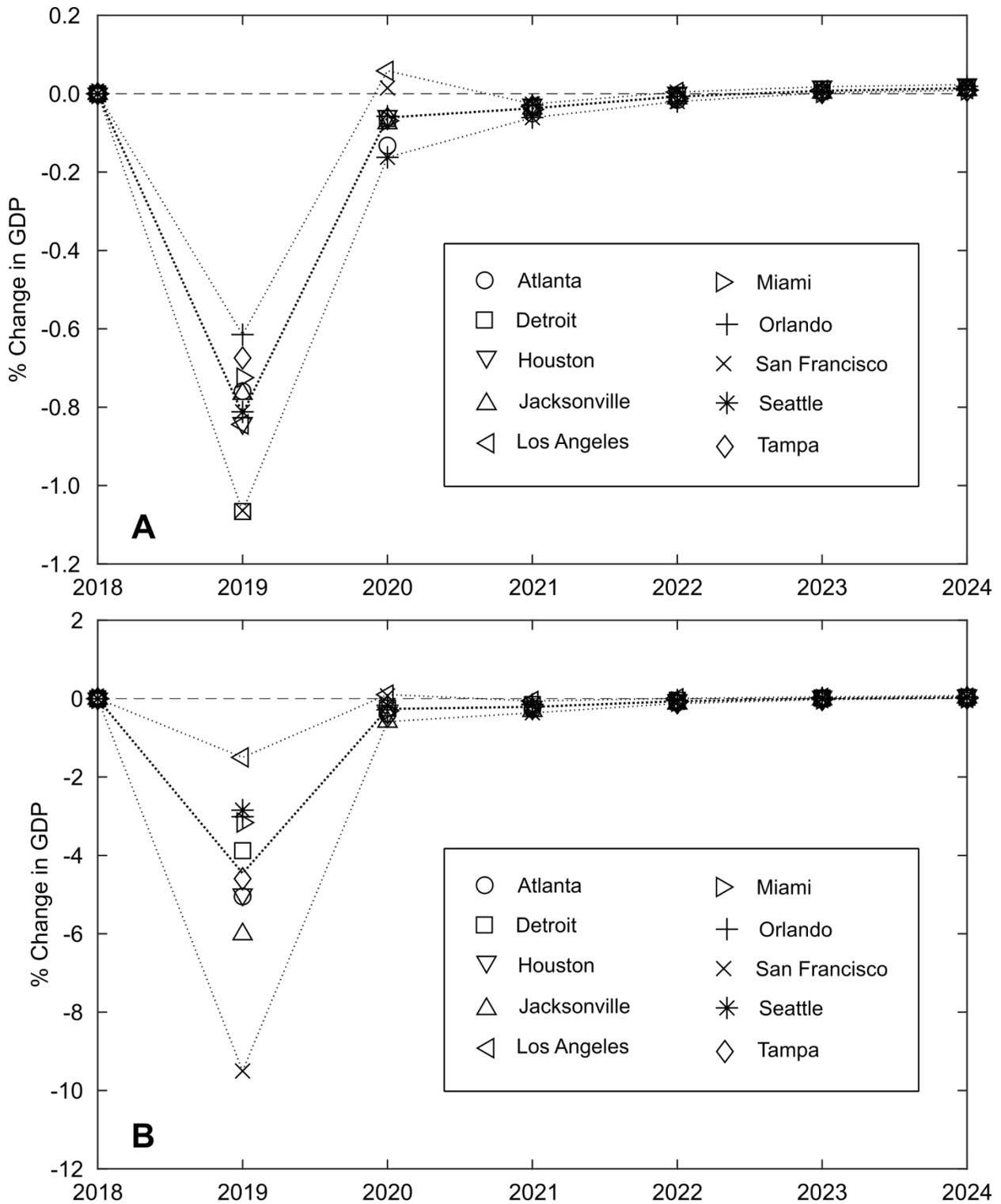

**Figure 2. Temporal performance of regional economies (as measured by GDP) in response to a 5% increase in travel costs that lasted one year (A) and to a 5% transportation network disruption (B). Dotted lines show to the lowest, mean, and highest values for each year. Dashed line corresponds to zero change in GDP from expected GDP in the absence of disruption.**

Figure 2 shows the temporal impact of a 5% increase in transportation costs on the GDPs for the baseline and test case scenarios. All simulated cities show significant impact on GDP in the year in which disruption occurs (2019) and the residual effect of that shock over the five subsequent years is relatively small. All of the cities recover to within 0.2% of their expected GDP (simulated GDP in the absence of a shock) by 2020 and have exceeded it by 2023. This still corresponds to a loss of, for example, $250,000 in Atlanta in the year following the shock. The profile of the initial impact and recovery is very similar for the test case, yet the magnitude is substantially greater, with disruption roughly an order of magnitude worse than the baseline case. Notice the difference in the y-axis of the Figure 2 graphs.

In Figure 3, GDP changes for the baseline scenario are displayed by the colored bars and those for the test scenario are shown with the transparent bars. For both scenarios, GDP progressively declines as a function of disruption severity $\rho$. At the same time, we note that changes in GDP due to travel time delays are significantly more consequential than their baseline counterparts. In this demonstration, for example, a random disruption of $\rho = 3\%$ of road segments in the San Francisco urban area results in $c(\rho) = 34\%$ transportation cost increase, which translates to 6.64% GDP decrease, significantly more than the baseline result of 0.64%. Not all cities show such disparate results between the two scenarios. For example, for Los Angeles (the 5th bar), a 1% roadway disruption increases the travel time by approximately 1% and therefore the GDP effects are the same (the transparent and the colored bars completely overlap).

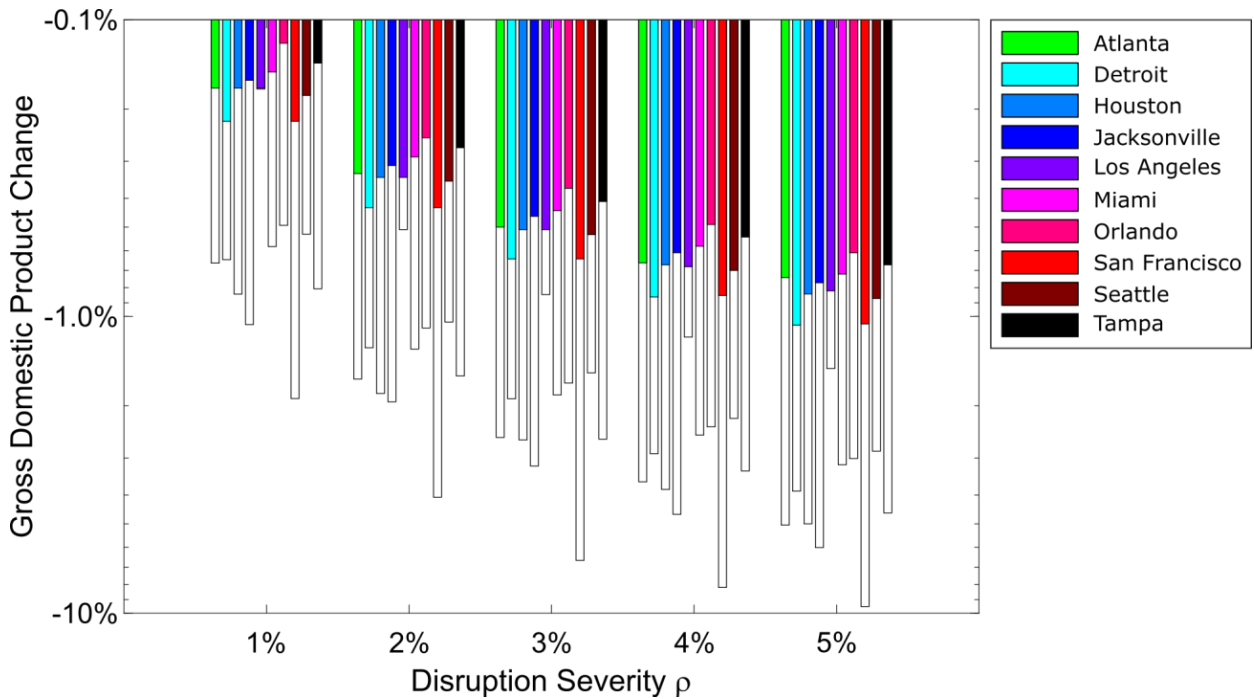

**Figure 3. Impact of disruption on GDP (vertical axis scale is logarithmic). The colored bars correspond to the baseline scenario with transportation cost increase being proportional to the disruption severity $\rho$ while the transparent bars present the case when transportation costs change per travel time increase in response to a disruption.**

To find how regional economies of different sizes are sensitive to unpredictable road disruption events we compare the percent changes in GDP, $\Delta g_i$, due to the random disruptions of $\rho = 5\%$ of road segments with the average GDP per capita values $G_i$ for each of the cities $i = 1,\ldots,10$ (Figure 4). For convenience, we show the mean values of $\langle G \rangle = \$73,422$ and $\langle \Delta g \rangle = -4.46\%$ on each of the axes with dashed lines. The worst response to disruption and a GDP decrease of 9.5% is found in San Francisco while the lowest GDP decrease of 1.5% is found in Los Angeles consistently with cost increases per Table 1. This clearly shows that the GDP changes under road disruptions do not correlate with the GDP per capita values under no disruption conditions.

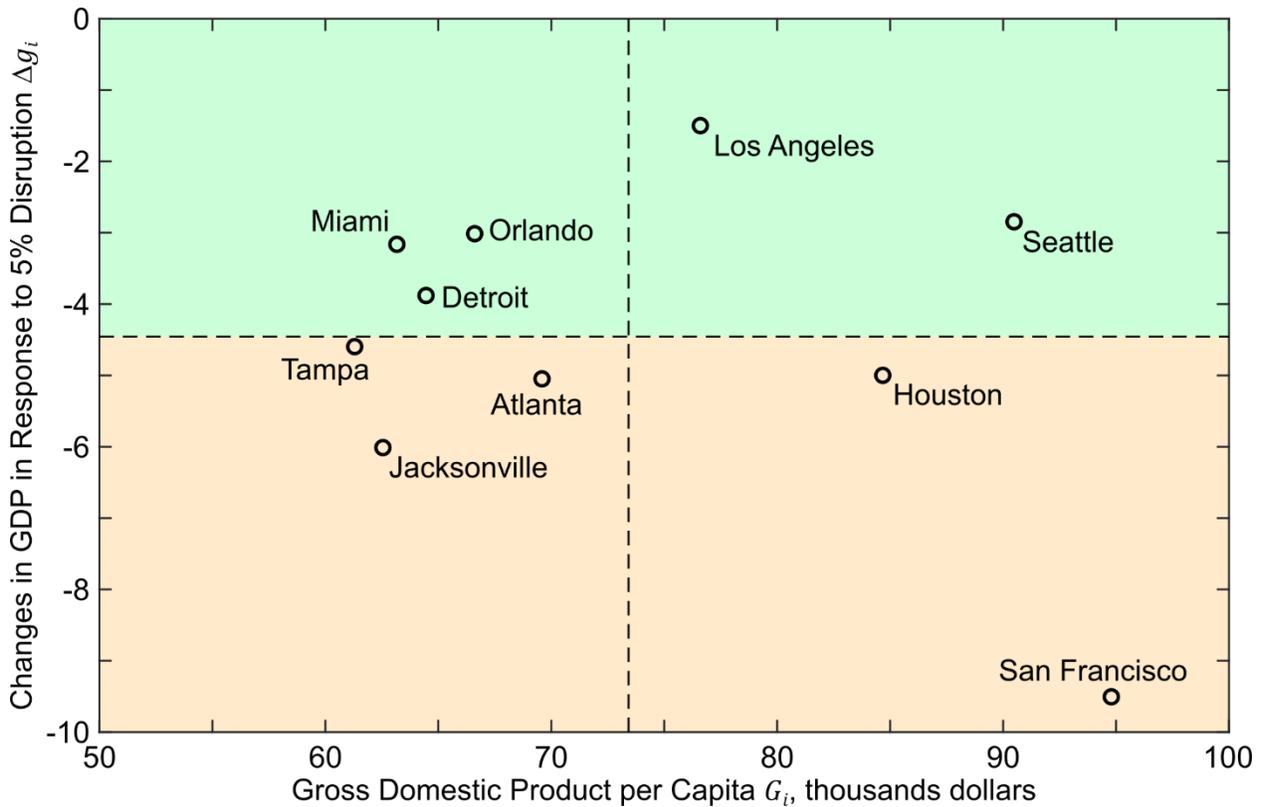

**Figure 4. Scatter plot of GDP per capita and change in GDP per capita in response to a 5% increase in transportation costs. The dashed lines indicate the means of the plotted values.**

## Discussion and Conclusions

Improving the ability of infrastructure systems to maintain functioning in the face of unexpected disruptions (i.e., resilience) is emerging as a high priority in infrastructure planning and an objective that has to be balanced against other system performance objectives, including efficiency and of transportation networks. (Ganin et al., 2017) showed that efficiency and resilience are not correlated in the case of 40 cities in the USA, but economic implications of planning for efficiency and neglecting resilience had not been addressed. It is intuitive that the implications of disruptive events are not isolated to the infrastructure that is directly impacted and yet, the extent that impacts can propagate through regional economies need to be quantified to motivate remedial action and inform priorities. Standard practices for how to incorporate resilience thinking into planning and management are broadly lacking but, as this research demonstrates, there are existing methods that are appropriate for creating and processing metrics of resilience. Here, the network model produces additional delay, which is designated as a metric of resilience and is compatible as an input to TranSight.

This research demonstrates that, in developing the workflow that can support resilience analysis and decision making, economic models must be paired with network analysis in order to best reflect the impact of disruption on economically important processes, such as commuters accessing their workplaces and firms moving their commodities to market. This work dispels the viability of the assumption that degree of disruption (i.e., percent of roadways disrupted) should be set as proportional to transportation cost increase. It follows that one important conclusion of this work is that losses associated with network disruptions may be an order of magnitude higher than the size of disruption itself, as evidenced in Figure 3. The model is limited however in that the calculated transportation cost increase was applied for a period of an entire year; a finer temporal resolution would be preferable. On the other hand, in reality, unexpected delays, can be expected to result in disproportionally higher cost increases as compared to expected delays. For example, traffic congestion does not disrupt businesses on a day to day basis because it can be planned for whereas unforeseen delay cannot be planned around and may cause disruptions that hurt businesses. Because Value of Reliability is not taken into account in the economic model used here, the economic impact of transportation disruption are conservative.

The finding that economic effects of road disruptions do not scale with the size of city economies (in terms of GDP per capita) and that not all of the regional economies in the study are equally affected by the adverse events merits future research to gain insight into what makes some cities more sensitive than others. A naïve expectation is that the larger the economy, the more sensitive it will be to transportation infrastructure disruption however, the results of this study point to the likelihood that more complex processes are at work. Similarly, cities recover from increased transportation costs differently, as shown in Figure 24, which is further evidence of their differing sensitivity to disruptions. We can speculate that sensitivity of an economy is dependent on the reaction of the transportation network to disruption as well as the dependence of the economy on its transportation network. The implication is that both region-specific economic and transportation models are necessary for resilience planning. Future work should explore the driving factors of uneven outcomes across cities, namely the reasons for transportation network sensitivities to disruption as well as local economy sensitivities to transportation failures and ability to recover. Additionally, future work can advance areas in which this study was limited and move the method toward practical application in transportation planning. For example, the topological attributes of road networks that yield more or less delay are not called into question in this research. Similarly, the mechanisms by which network disruptions cause economic impact in the results are not investigated. Efforts to enhance the resilience of road networks to disruption and/or the regional economy to lack of road network resilience will need to study the outcomes of the models in detail. (Rose, 2017) gives a comprehensive overview of the challenges faced in accounting for the economic impacts of disruption including that of translating damaged public infrastructure into broader economic loses. Other limitations include that the network models that underpin delays do not currently incorporate public transportation options, a significant mode in some cities, most notably New York City. Similarly, the traffic model is intentionally abstract and simple and only uses publicly available data sets and economic variables have a time step of one year.

A key motivation of this research is to highlight that efficiency, risk reduction and resilience are not the same objectives and that paring process models with economic simulation should stimulate greater attention to the differences. Although the outcome of disruptive events is delay, and sometimes much more pronounced delay than routine congestion, they cannot be prevented by efficiency improvement or risk reduction; planning for resilience is functionally different than planning for efficiency (Ganin et al., 2017) and risk reduction can have limited success for uncertain events. For example, in flow networks, such as transportation systems, efficiency may be achieved by having sufficient roadway capacity, while resilience can result from the availability of, potentially not very efficient, alternative routes, which would ensure graceful performance degradation under disruption as opposed to a complete collapse. As this paper demonstrates, the consequence of neglecting to plan for disruptive events from a resilience perspective is disproportionate impact to both the primary system and the connected, dependent economy.

Ganin, A.A., Kitsak, M., Marchese, D., Keisler, J.M., Seager, T., Linkov, I., 2017. Resilience and efficiency in transportation networks. Science Advances 3, e1701079. https://doi.org/10.1126/sciadv.1701079

Ganin, A.A., Mersky, A.C., Jin, A.S., Kitsak, M., Keisler, J.M., Linkov, I., 2019. Resilience in Intelligent Transportation Systems (ITS). Transportation Research Part C: Emerging Technologies 100, 318–329. https://doi.org/10.1016/j.trc.2019.01.014

Hallegatte, S., Rentschler, J., Rozenberg, J., 2019. Lifelines: the resilient infrastructure opportunity, Sustainable infrastructure series. World Bank Group, Washington.

Ham, H., Kim, T.J., Boyce, D., 2005. Assessment of economic impacts from unexpected events with an interregional commodity flow and multimodal transportation network model. Transportation Research Part A: Policy and Practice 39, 849–860. https://doi.org/10.1016/j.tra.2005.02.006

Highway Capacity Manual, 5th Edition, 2010. Transportation Research Board, Washington, D.C.

Lam, T.C., Small, K.A., 2001. The value of time and reliability: measurement from a value pricing experiment. Transportation Research Part E: Logistics and Transportation Review 37, 231–251. https://doi.org/10.1016/S1366-5545(00)00016-8

Mackie, P.J., Wardman, M., Fowkes, A.S., Whelan, G., Nellthorp, J., Bates, J., 2003. Values of Travel Time Savings UK (No. 567). The University of Leeds, Leeds, UK.

Mattsson, L.-G., Jenelius, E., 2015. Vulnerability and resilience of transport systems – A discussion of recent research. Transportation Research Part A: Policy and Practice 81, 16–34. https://doi.org/10.1016/j.tra.2015.06.002

Model Equations, 2017. Regional Economic Models Inc., Amherst, MA.

Obama, B., 2013. Presidential Policy Directive 21 - Critical Infrastructure Security and Resilience. U.S. Office of the Press Secretary.

Pelling, M., Özerdem, A., Barakat, S., 2002. The macro-economic impact of disasters. Progress in Development Studies 2, 283–305. https://doi.org/10.1191/1464993402ps042ra

Rose, A.Z., 2017. Defining and measuring economic resilience from a societal, environmental and security perspective, Integrated disaster risk management. Springer, Singapore.

Schrank, D., Eisele, B., Lomax, T., Bak, J., 2015. 2015 Urban Mobility Scorecard. Texas A&M Transportation Institute and INRIX, College Station, TX.

Tatano, H., Tsuchiya, S., 2008. A framework for economic loss estimation due to seismic transportation network disruption: a spatial computable general equilibrium approach. Nat Hazards 44, 253–265. https://doi.org/10.1007/s11069-007-9151-0

TranSight 4.2 User Guide, 2018. Regional Economic Models Inc., Amherst, MA.
13